\documentclass[%
 reprint,
superscriptaddress,
 amsmath,amssymb,
 aps,
]{revtex4-1}

\usepackage{graphicx}
\usepackage{ulem}
\usepackage{dcolumn}
\usepackage{bm}
\usepackage{color}
\usepackage{amsmath}
\usepackage{hyperref}
\usepackage{xcolor}
\usepackage{url}
\usepackage{hyperref}
\hypersetup{
    colorlinks=true,allcolors=blue,}
\begin{document}

\preprint{APS/123-QED}

\title{Low power In Memory Computation with Reciprocal Ferromagnet/Topological Insulator Heterostructures}
\author{Hamed Vakili}
\email{hv8rf@virginia.edu}
	\affiliation{
		Department of Physics, University of Virginia, Charlottesville, VA 22904, USA\looseness=-1}
\author{Samiran Ganguly}%
	\affiliation{
		Department of Electrical and Computer Engineering, University of Virginia, Charlottesville, VA 22904, USA\looseness=-1}
        \affiliation{
		Department of Electrical and Computer Engineering, Virginia Commonwealth University, Richmond, VA 23284, USA\looseness=-1}
\author{George J. de Coster}
	\affiliation{
		DEVCOM Army Research Laboratory, 2800 Powder Mill Rd, Adelphi, MD, 20783, USA\looseness=-1}
\author{Mahesh R. Neupane}
	\affiliation{
		DEVCOM Army Research Laboratory, 2800 Powder Mill Rd, Adelphi, MD, 20783, USA\looseness=-1}
	\affiliation{
		Materials Science and Engineering Program, University of California, Riverside, CA, 92521, USA\looseness=-1}
\author{Avik W. Ghosh}
	\affiliation{
	    Department of Physics, University of Virginia, Charlottesville, VA 22904, USA\looseness=-1}
	\affiliation{
			Department of Electrical and Computer Engineering, University of Virginia, Charlottesville, VA 22904, USA\looseness=-1}

\begin{abstract}
The surface state of a 3D topological insulator (3DTI) is a spin-momentum locked conductive state, whose large spin hall angle can be used for the energy-efficient spin orbit torque based switching of an overlying ferromagnet (FM). Conversely, the gated switching of the magnetization of a separate FM in or out of the TI surface plane, can turn on and off the TI surface current. By exploiting this reciprocal behaviour, we can use two FM/3DTI heterostructures to design an {integrated} 1-Transistor 1-magnetic tunnel junction random access memory unit (1T1MTJ RAM) for an ultra low power Processing-in-Memory (PiM) architecture. Our calculation involves combining the Fokker-Planck equation with the Non-equilibrium Green Function (NEGF) based flow of conduction electrons and Landau-Lifshitz-Gilbert (LLG) based dynamics of magnetization. Our combined approach allows us to connect device performance metrics with underlying material parameters, which can guide proposed experimental and fabrication efforts.
\end{abstract}

\maketitle

In-memory computing or Processing-in-Memory (PiM) \cite{PiM,pim2} is an important emerging architectural design that reduces data movement between the memory and the processor. PiM operates by performing simple intermediate steps along a long chain of compute processes within the memory array itself, as far as possible. The memory layout in a typical PiM architecture is in the form of a grid. Each row and column of the grid is driven by selectors, which enable the cells for a read or write operation. A sense amplifier reads an entire row of the memory cell by comparing its state against a known reference voltage, current, or charge \cite{sttPiM}. While such a local computing paradigm leads to a significantly reduced footprint, this however, needs to be traded off against  material integration complexity as well as overall switching costs.

Recent experiments on spin-orbit torque (SOT) based switching \cite{FMTI,mellnik_spin-transfer_2014,wang_room_2017,han_room-temperature_2017,han_topological_2021,wu_magnetic_2021} in FM/3DTI heterostructures suggest TIs as a promising alternative to heavy metal underlayers, because of their higher spin Hall angle. Conversely, the ability of a FM to turn current ON or OFF in a TI by breaking {time reversal} symmetry through its orientation \cite{semenov_electrically_2012,taniyama_electric-field_2015}
offers an option for a gate tunable selector, atop the intrinsic energy efficiency of a gate-tunable  bandgap \cite{ghoshacs,grpnj1}. This brings up intriguing possibilities of using the reciprocal interactions between a FM and a TI to realize energy-efficient device configurations.

\begin{figure}[h]
    \centering
    \includegraphics[width=1.0\linewidth]{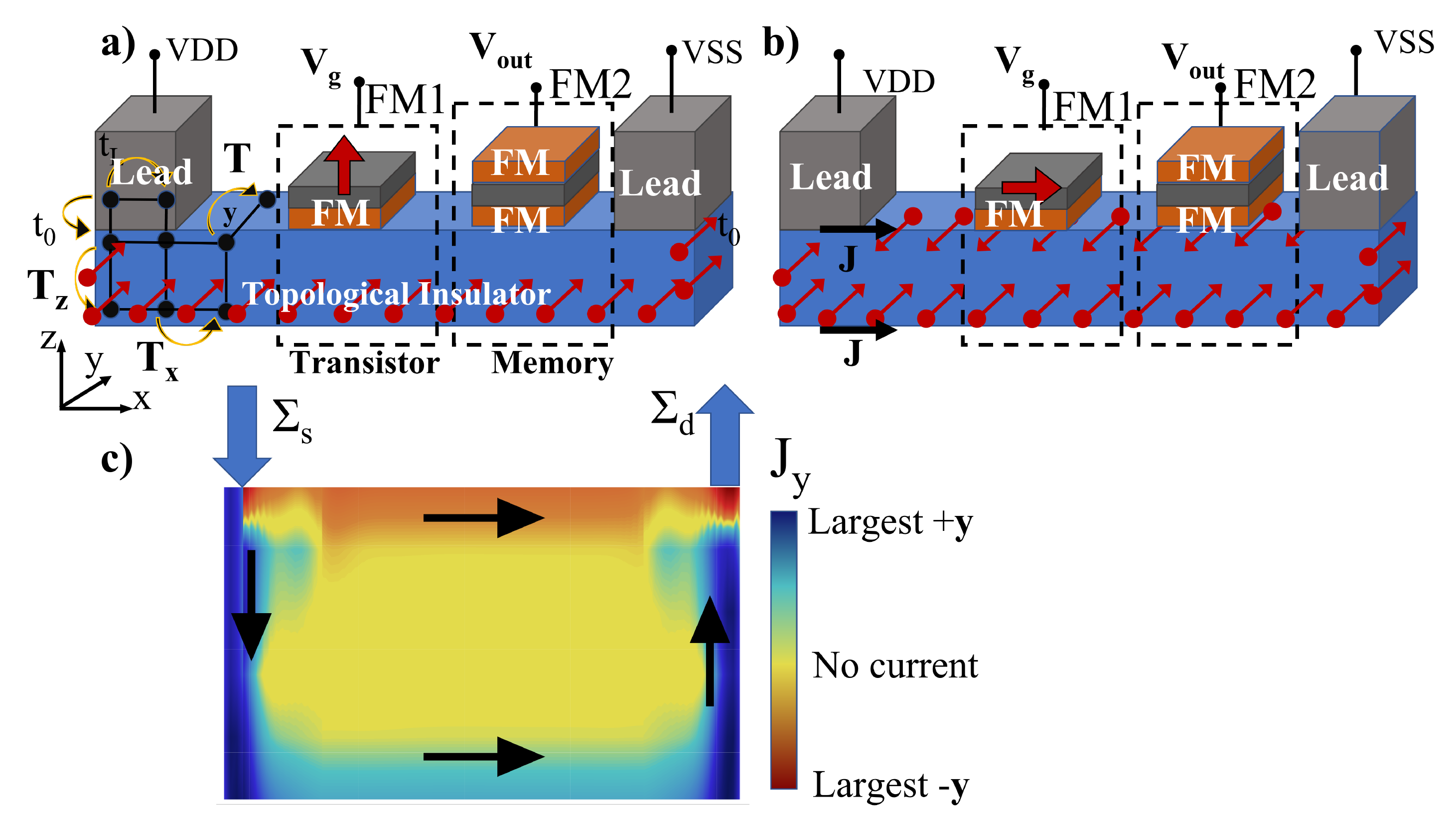}
    \caption{a) Schematic 1T1MTJ cell using FM/3DTI heterostructures {sitting well within a spin scattering length}. The first ferromagnet (FM1) is the selector unit while the second (FM2) is the memory unit. In the off state, \textcolor{black}{FM1 is oriented perpendicular to the TI plane to open a gap in its top surface states (Fig.~\ref{fig:band})}. \textcolor{black}{The large red arrow shows the FM polarization and the spheres with small arrows show the spin current }b) In the ON state, ungapped metallic states on the top surface and surface current are restored by rotating FM1 into the TI plane. c) Simulated current flows on top and bottom layers,  connected by current flowing \textcolor{black}{along the ungapped side walls}. \textcolor{black}{Note that the oxide does not need to be between the FM1 and the TI surface.}}
    \label{fig:Device}
\end{figure}
In this paper we present a 1-Transistor 1-magnetic tunnel junction  random access memory unit (1T1MTJ-RAM),  that can  function as a potentially compact, energy efficient building block of a PiM (Fig.~\ref{fig:Device}). The proposed device has two FM/3DTI heterostructures, one functioning as a row-column selector switch, the other as a nonvolatile memory unit. Switching in the second  MTJ memory unit (FM2) is based on conventional SOT, with the required spin current at the FM2-3DTI interface provided by the spin momentum locking at the 3DTI surface. The FM that acts as the selector unit (FM1) needs to be electrically switched from out of plane to in plane and back. There are a number of mechanisms to achieve this switching, such as a gated piezoelectric clamped on a magnetostrictive material sitting on the TI  \cite{vaz_intrinsic_2015,trassin_low_2016,DFTstrain,de_ranieri_piezoelectric_2013,Piezo,VCU_strain}, or voltage Controlled Magnetic Anisotropy (VCMA) at the FM/3DTI \cite{vcma} to control the polarization of the FM1. Another mechanism is to tune the interface induced anisotropy of a FM/3DTI by changing the free energy of the heterostructure using an applied gate voltage \cite{kiwook,kiwook2}. \textcolor{black}{Although the electrical switching of ferromagnets in FM/3DTI remains challenging, this proposal could encourage more experimental efforts in this direction.}
 
 One of the known challenges for the FM1-3DTI is a low On/Off ratio compared to competing CMOS technologies. In this paper we show that as a selector for low power PiM, the On/Off ratio does not need to be very high.
In fact, our proposed device is naturally suited for compact PiM designs, as it directly incorporates a selection transistor in the first FM1-3DTI combination. Using an enhanced sense-amplifier for each column with programmable sensing thresholds, it is possible to implement basic Boolean operations (AND, OR, XOR, Majority, and their complements). We describe a possible scheme of building such a PiM towards the end.

\section{Results}

\textbf{Switching Mechanism} The SOT is calculated from the in and out-of plane components of the polarization current applied to the FM2 $\mathbf{J_p} = \mathbf{J}_{in}+i\mathbf{J}_{out}$, obtained by adding phenomenological scattering terms within FM2 of thickness $t_{FM}$ \cite{torque,vaezitoque,vaezi2} on top of the  NEGF calculated (Eq.~\ref{spincurrentdensity}) spin current $\mathbf{J_s}$ in the TI
\begin{subequations}
\begin{eqnarray}
    \mathbf{J}_{p} &=& \frac{p\mathbf{J_s}L^2}{t_{FM}}\bigg(\frac{1}{\lambda_{\phi}^2}-\frac{i}{\lambda_j^2} \bigg)\frac{\sinh{t_{FM}/L}}{\cosh{t_{FM}/L}}\label{FMcurrent}\\
    {\mathbf{\tau}}_{SOT}&=&\mathbf{m\times (J_p\times m) - \alpha\; m\times J_p}
    \label{eqjp}
\end{eqnarray}
\end{subequations}
where $\alpha$ is the Gilbert damping, $\mathbf{m}$ is the FM magnetization vector, and the net scattering length $L = (1/\lambda_{sr}^2+1/\lambda_{\phi}^2-i/\lambda_{j}^2)^{-1/2}$. $\lambda_{sr},\lambda_{\phi},\lambda_{j}$ are spin relaxation, precessional and decoherence length of the FM, while p is the interface dependent spin transfer efficiency from TI to FM2 (assumed 0.5). Note that due to spin precession in FM2, an out of plane component arises in the SOT, denoted by the imaginary part of $\tau_{SOT}$.

\begin{figure}
    \centering
    \includegraphics[width=1.0\linewidth]{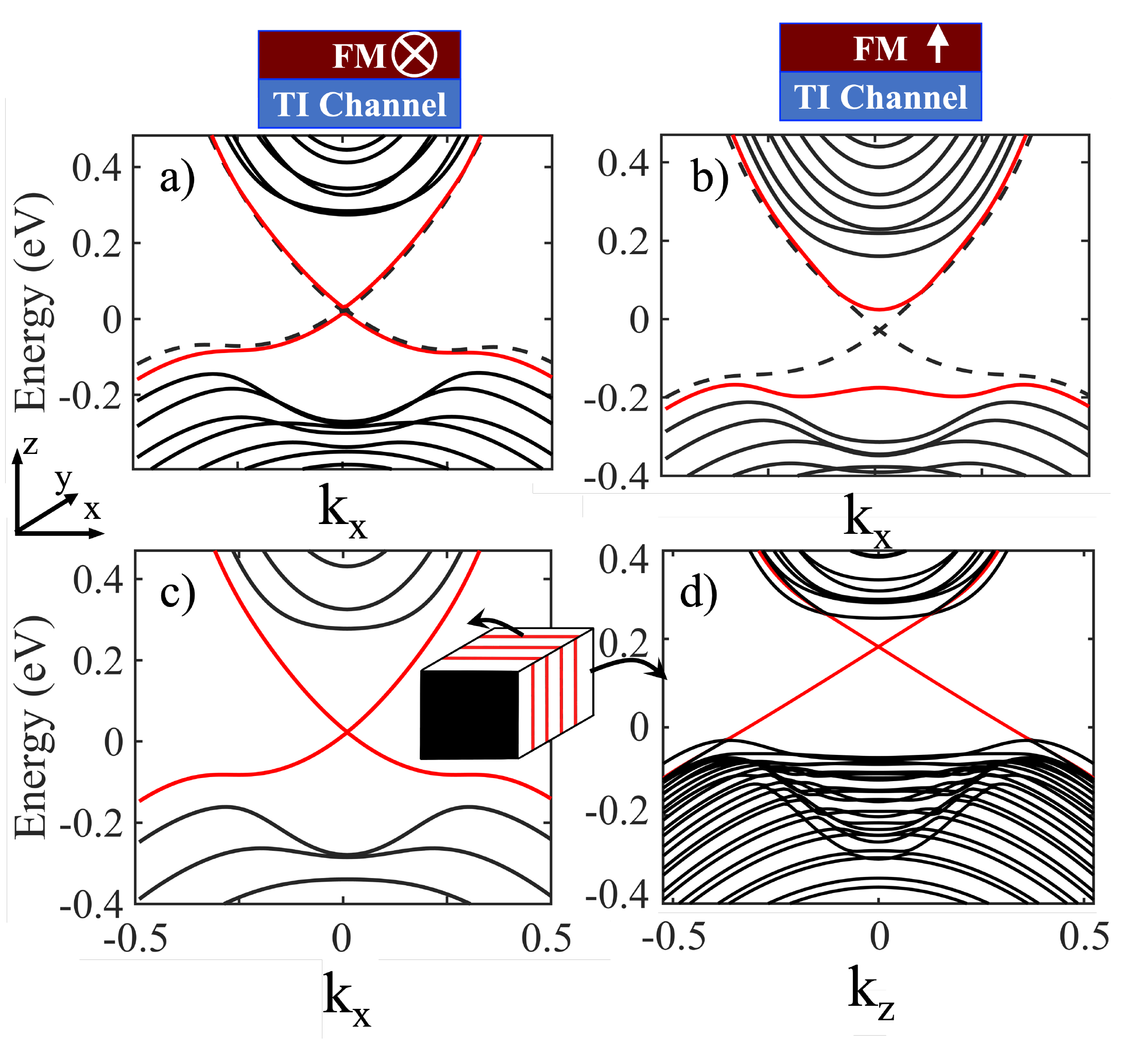}
    \caption{Band structure of the 3DTI at various points. Red lines in (a-c) emphasize the top surface states and in d, the side surfaces. \textcolor{black}{Dashed lines show the bottom surface states} a) FM/3DTI bandstructure with in plane magnet (in y direction) b) Magnet in z direction with on site voltage energy at the top at -0.1 and bottom at -0.05 V. c) Bandstructure of the \textcolor{black}{pristine (magnet free)} 3DTI channel discretized in the z direction (top to bottom) d) {3DTI side surface (with magnet) bands discretized in the transport direction, show ungapped states even when the FM1's exchange coupling extends to a few top TI layers.}}
    \label{fig:band}
\end{figure}

A gate tunable anisotropy (through strain or VCMA) switches FM1 from out of plane to in plane, which changes the magnetic exchange energy. \textcolor{black}{Note that when the FM1 is in plane, it does not need to be polarized in any specific direction as long as out plane component is negligible.} The bandstructure of the 3DTI below FM1 is modified accordingly, shifting locally relative to the Fermi energy (Fig.~\ref{fig:band}). We  shift the bottom surface 0.5 times the top surface under the applied gate voltage $V_g$, \textcolor{black}{and assume the magnet covers the full TI width laterally}. The corresponding NEGF calculated torque $\tau_{SOT}$ with phenomenological corrections (Eq.~\ref{eqjp}) is then fed into the Landau-Lifshitz-Gilbert (LLG) equation \cite{SLONCZEWSKI1996L1}  self-consistently and quasi-statically, assuming the electron transit speed is orders of magnitude faster than the magnet's resonant frequency. 

The voltage control of the FM1 anisotropy can arise from VCMA effect driven by a vertical electric field across an oxide(MgO)-FM stack, $\Delta K = {\xi V}/{t_{ox}t_{FM}}$, with VCMA coefficient $\xi$ and oxide thickness $t_{ox}$ and $\Delta K$ the change in the anisotropy. Alternatively, if we use strain as a switching mechanism for FM1, the applied effective field can be added to $\mathbf{H}$ as $\zeta \mathbf{(u.m)u}$ where $\zeta$ is the strain coefficient that needs to cancel out the effective anisotropy field $\mathbf{H_K}$. u is the strain direction which needs to be along z (perpendicular to the TI) in this case. For the VCMA mechanism, assuming a thickness of 1 nm for the capped oxide and FM1 layers, a $\Delta K/V = 33-100 kJ/m^3$ (\textcolor{black}{change in anisotropy volume density}) for CoFeB of 1 nm thickness has been reported \cite{cofeb50,cofeb,rana_towards_2019}, whereas by doping FM/oxide interface a larger $\Delta K/V$  has been achieved as well \cite{vcmareview,vcmaadvance}. \textcolor{black}{The effect of oxide on the FM1/3DTI interface is ignored here as it is on the other side of the FM1 (not sandwiched between FM1 and 3DTI), and the effect of the FM1 and FM2 on the bandstructure of the 3DTI is assumed to be limited to the magnetic exchange.} For the strain mechanism, $\Delta K/V$ of 200-300 $kJ/m^3$ has been reported. For both the VCMA and strain, the free energy change from the applied gate $V_g$ can potentially lower the voltage requirements. The required voltage to change the anisotropy of a reliable FM1 with thermal barrier $\Delta_T =40$ of size $40\times40\times 1$ nm$^3$ would be 0.3-1 V. Although this is relatively large compared to the applied bias, since there is negligible current in the FM1 heterostructure {(we assume a thin insulating buffer)}, the energy consumption is still small ($\sim$ 100-300 aJ). \textcolor{black}{Most of the energy consumption comes from the current dissipation in the 3DTI. Using the material parameters reported for room temperature 3DTI switching of ferromagnet \cite{dc_room-temperature_2018}, the energy consumption from the current in the 3DTI is estimated to be around 10 fJ.}

\textcolor{black}{We set up a tight binding Hamiltonian that describes the 3DTI, the Zeeman energy term $H_Z$ (magnetic exchange) originating from the FM/3DTI exchange and the on-site energy term $H_V$ from the applied gate voltage at FM1, with parameters fitted to {\it{ab-initio}} calculations (details in the Method section).}

Figs~\ref{fig:band}a and b show the bandstructure of the 3DTI at the location of the FM1, for the On (magnet in plane of TI) and Off (magnet out of plane) states respectively. For the On state (a), we see a shift in the Dirac point away from the $\Gamma$ point, relative to the pristine (magnet-free) TI states (Fig.~\ref{fig:band}c). For the Off state (b) however, the degeneracy is lifted \textcolor{black}{and a corresponding energy gap is created for the top surface states (red solid lines), while at the same time keeping the bottom surface states (dashed lines) intact}. We also present the side surface states (Fig.2.~d), and as expected (Fig.~\ref{fig:Device}b) they stay gapless for both On and Off states, as the FM only affects part of the top surface. This would mean that for an electrical transistor based on FM/3DTI, {the side current would play a crucial role. However, using FM1/TI as a selector as proposed}, we avoid dealing with the side current as \textcolor{black}{only the top surface states can apply any appreciable spin orbit torque to FM2}.

\textbf{Reliability} For a competitive transistor-memory device based on the FM/3DTI stack, we outline requirements in this paper. The first is the On/Off ratio $\beta$. For deterministic switching of an in-plane ferromagnet with a given thermal barrier $\Delta_T$, the critical spin current $I_c = ({4q\alpha k_B T\Delta_T}/{\hbar})(1+H^*_D/2H_K)$ \cite{sun} with $H^*_D$ being the effective demagnetization field which we take to be twice $H_K$.
We also need to make sure that when FM1 is out-of-plane in the Off state, the surface leakage current would be small enough to not accidentally switch FM2. The probability $P$ of switching FM2 can be approximated using the Fokker-Planck equation\cite{Lifokker,butlerfokker,fokkerinplane}
\begin{eqnarray}
    P &\approx& 1-\Delta_T\frac{\iota}{\iota+1}e^{-2\iota\tau},\\
    \tau &\equiv & \frac{\alpha \gamma \mu_0 H_K}{1+\alpha^2}t\nonumber
    \label{Prob}
\end{eqnarray}
When the TI is not gapped (FM1 in plane, $P \approx 1$), the write error $WER \approx 1-P$, while when locally gapped (FM1 out of plane, $P \approx 0$), the error is $WER \approx P$. For simplicity we take {our target WER} in both cases to be identical.
For thermally assisted switching (Off) we can solve the Fokker Planck equation, or use the empirical equation $P = \exp{(-tf_0 \exp{(-\Delta_T (-\iota)^c))}}$.
where t is the switching time, and  $\iota=I_s/I_c-1$ is the spin current overdrive. c is an empirical constant, assumed 1 for in-plane FM2, 2 for out-of-plane, and in between for imperfections. 
\begin{figure}
    \centering
    \includegraphics[width=1\linewidth]{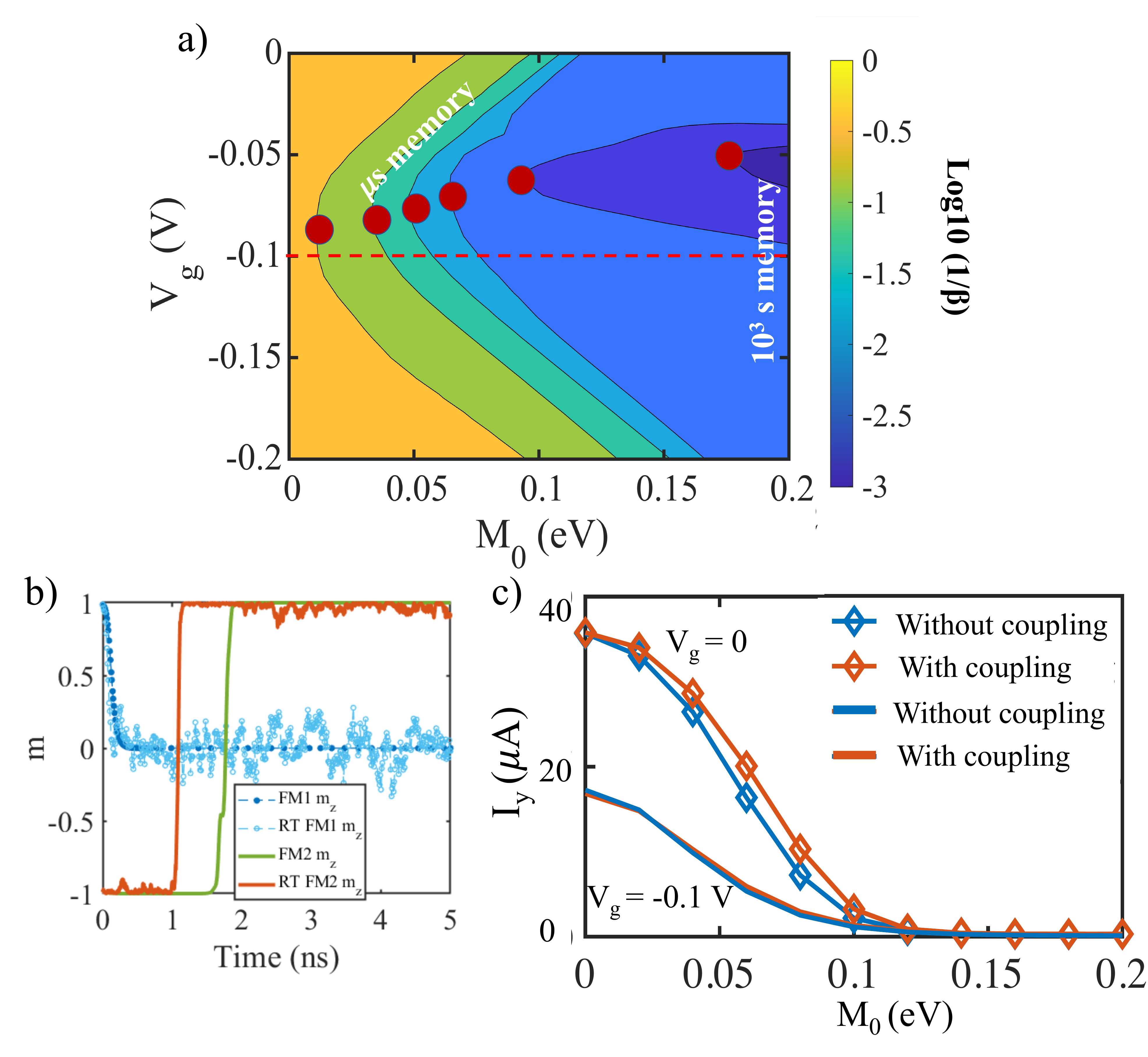}
    \caption{a) The colorplot shows the logarithm of the On/Off ratio $\beta$ of non-equilibrium spin current $J_s$ at the FM2/3DTI interface with respect to the Zeeman energy coefficient $M_0$ and applied gate bias $V_g$. The contours show different memory regimes in which the device can operate, ranging from cache memory (operating in $\mu s$ up to longer term memory, like a computer RAM $10^3$ s). The dashed red line is the $V_g$ equal to the applied bias VDD. As explained in the text, the red circles show the minimum required $M_0$ for each contour. b) Figure shows the 0 K and 300 K LLG simulation of the \textcolor{black}{FM2 magnetization dynamics}. Applied voltage bias and $V_g$ are 100 $mV$. The Zeeman exchange energy for FM2 is taken to be 10 $meV$. $\lambda_{sr},\lambda_{\phi},\lambda_{j}$ are taken to be 1, 10, 1 $nm$ respectively. Source, drain, FM1 and FM2 lengths are 20 nm each. 3DTI thickness is 5 nm (6 layers) and magnet thickness t is 2 nm. \textcolor{black}{The total length of the channel is 140 nm}. Gilbert damping $\alpha$ is 0.1. c) \textcolor{black}{The current flow with and without electric coupling between FM1 and 3DTI at $V_g =$ 0 (diamond) and -0.1 V. The coupling between FM1 and 3DTI top surface is taken to be 0.3 eV which corresponds to effective mass of $m* = 0.48 m_e$ where $m_e$ is electron mass}.\cite{sankey_measurement_2008}} 
    \label{fig:OnOff}
\end{figure}

The Zeeman term $H_Z = M_0 \mathbf{S}.\mathbf{\sigma}$ which breaks  time reversal symmetry to open a surface gap must be large enough to allow a small  WER for accidental switching. For the Off state, t should be $\sim\mu s$ for cache memory and years for long term memory. Eq.~\ref{Prob} connects WER to material parameters, and Fig.~3.a shows that depending on $V_g$ and $M_0$, the device can be used in different regimes. For  long term storage, we assume $\Delta_T = 40$, $WER = 10^{-7},  c=1.2$ \cite{switchingexp,switchingexp2} and attempt frequency $f_0 = 1~GHz $ \cite{attempt}.  When the FM1 is in plane, the required ON $I_s/I_c$ is $\sim0.2$. For short term memory, the off state will need $I_s/I_c$ to be $< 0.2$ to avoid accidentally switching FM2 with the above WER. This means that a modest On/Off ratio of $10$ would make the device work. For longer memory applications (hours to days), a larger On/Off ratio (Fig.~\ref{fig:OnOff}a), $\log_{10} \beta >3$ is needed. \textcolor{black}{$M_0$ of up to 100 meV has been reported in $MnBi_2Se_4/Bi_2Se_3$ heterostructure\cite{hirahara_large-gap_2017,PhysRevMaterials.5.124204} which is in the range for up to $10^2$ cache memory.}

In Figs.~\ref{fig:band}c,d we see an energy offset between the TI side and top surface bands, implying different surface conductivities at any given energy. To achieve minimum required magnetic exchange $M_0$, the applied gate voltage should be approximately equal to the applied bias $VDD$, $V_g \approx$ VDD. This matches Fig.~\ref{fig:OnOff}a (VDD $= -0.1$ eV, VDD $= 0$ eV), where the minimum $M_0$ required for short term ($\sim \mu$s) memory occurs when $V_g$ is near the midpoint (red dashed line). This implies that without a magnetization, when the chemical potential is tuned to the top or side Dirac point to reduce its conductivity, the other surface will retain a large density of states (DOS), providing a shunting conduction channel between top and bottom. For long term ($10^3$ s) memory, a higher On/Off ratio is needed which requires a bigger bandgap, hence a larger $M_0$. Due to the sizable gap opening for long term memory, the critical $V_g$s for the minimum $M_0$ (red circles) deviate from the midpoint. \textcolor{black}{In Fig.~\ref{fig:OnOff}c, we performed the simulation with and without any electric coupling between FM1 and 3DTI, which shows that at $V_g = 0$, the required $M_0$ is only increased by 5-10 percent. At higher $V_g$, there is no noticeable difference.}

Note that the On/Off ratio would be larger if we used a simplified 2D Hamiltonian that only considers surface states. This is because in a realistic TI structure, a significant amount of the current shunts into the bulk of the TI stack, which our 3D geometry \textcolor{black}{naturally} takes into account \textcolor{black}{(Fig.~\ref{fig:Device}c)}. The shunting allows the surface current to go around the gapped region, reducing the On/Off.
To have a faster working memory, using a perpendicular ferromagnet as FM2 is preferred, as the switching mechanism would be determined by a field like torque which is faster than an anti-damping torque. However, for a perpendicular ferromagnet an assisting external field is required. This assisting field can originate from the stray field of a capping magnetic layer or the exchange bias of a coupled antiferromagnet-ferromagnet stack  \cite{fieldsot,kong_spinorbit_2019,zheng_field-free_2021}. 
\begin{figure}
    \centering
    \includegraphics[width=\linewidth]{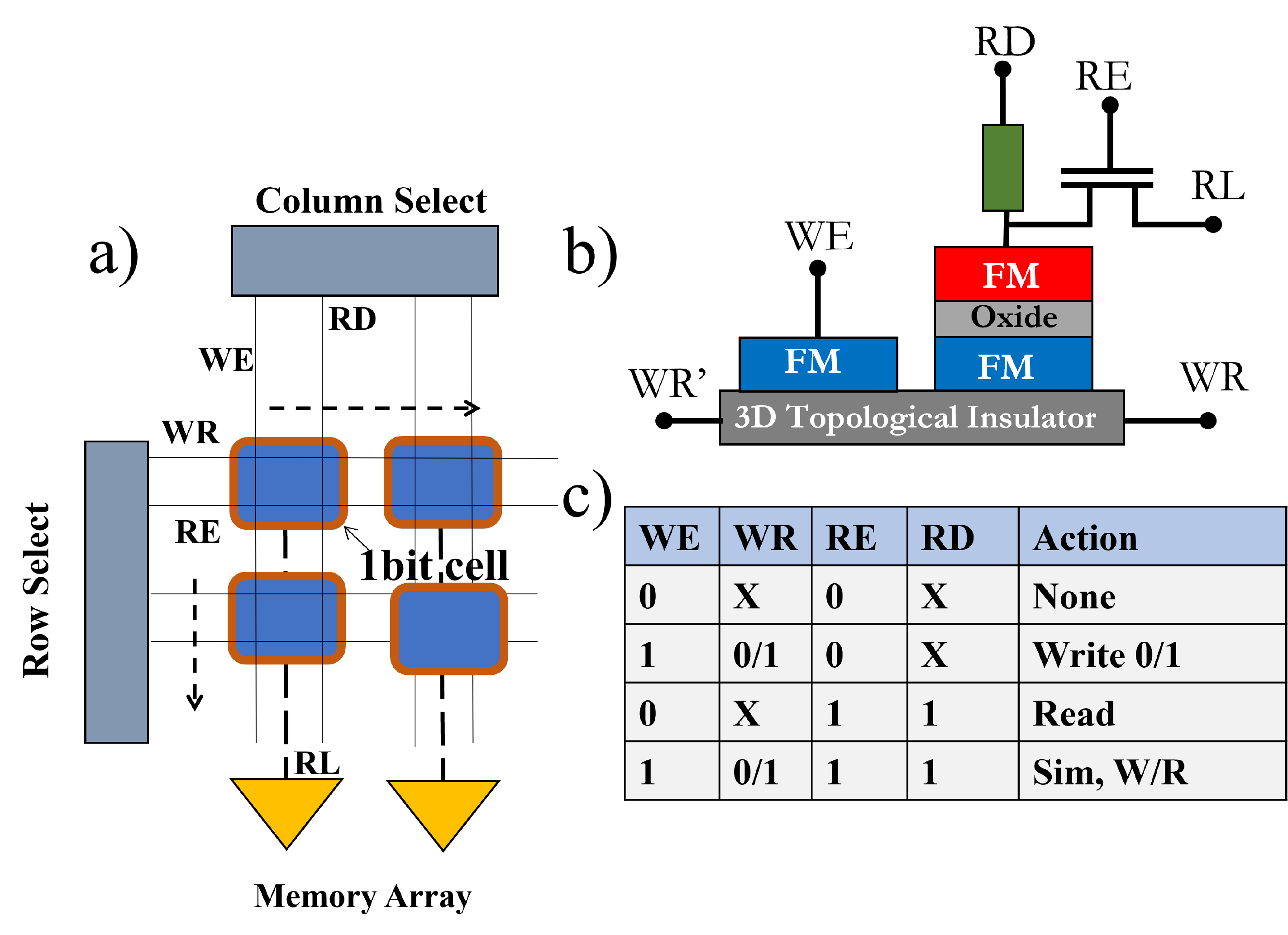}
    \caption{a. Memory array for the presented cell b. The cell and its position in the array indicated. c. The truth table for the array operations.}
    \label{fig:pim}
\end{figure}

{\textbf{Proposal for a Processor-in-Memory (PiM) design.}} In Fig. \ref{fig:pim}a. we show a possible approach for performing in-memory compute functions using an array built out of the presented memory cell. The relation between the voltage lines in Fig.~\ref{fig:pim} and \ref{fig:Device} are as follows: $WR:$ VDD, $WR':$ VSS, $WE:$ Vg, $RL: V_{out}$. 
The cell (Fig. \ref{fig:pim}b) contains a selection transistor, which acts as a write enable $WE$ signal. The current flowing in the TI acts as the write signal, with the two ends designated as $WR$ and $WR'$. Since a spintronic memory cell requires bipolar currents for programming, we can either use a bipolar current generating selector to drive the $WR$ signal with $WR'$ grounded, or else reverse the polarities of $WR, WR'$ pairs between $0/1$ and $1/0$ respectively. 

The read is performed at the reader MTJ with a voltage divider, which changes the output voltage at the read line ($RL$), shared column-wise over the array cells. The reader is charged through the $RD$ line. We include one transistor in the cell for a read enable functionality ($RE$) that  provides the load resistance for adequate current to the read line per cell depending on the voltage at the divider, critical for the read and PiM functionalities as we describe next. We choose to arrange the read to be done row-wise, whereas the write is done column-wise. This is by no means a necessary condition for designing this architecture, but simply one of the choices we make that allows us the option of simultaneous read and write, if necessary. The read is performed by sensing the current in the read line and comparing with a reference current in the sense amplifier (SA), which then reports the value stored at a specific cell addressed via $RD$ and $RE$ signals. The truth table of these operations is shown in Fig.~\ref{fig:pim}c.

The PiM functions are also arranged within the SA over each column. We can build in reference currents that enable Boolean operations over the whole array, by enabling multiple rows at a time. Consider a two-bit $AND$ operation over any two rows. In this case, the two specific rows are enabled using $RE$ and all the $RD$ over the rows are enabled as well. Each of the SA reference currents for the AND operation is set to $1.8I_R$ where $I_R$ is the read current of a cell when it is storing `1'. In this case both of the cells being read simultaneously by a single SA will have to be `1' to trigger the SA to report `1', in all other combinations it will report `0'. Similarly for the OR case, the reference current can be set as $0.8I_R$ to trigger `1' from SA for even one of the cells being `1'. An XOR gate can be implemented by a two threshold SA where the thresholds are $0.8I_R$ and $1.8I_R$ and it is configured to report `1' when reading values in between these two references. The complement functions are easily implemented by using an additional inverter. All these functionalities are built within a single SA and as per the requirements of the computation the SA can be configured to perform a given operation. 

\textcolor{black}{
{\textbf{Comparison With Existing Technology}}
The in-memory computation is also possible to be performed with existing technologies such as SRAM and DRAM. Note that in order for our proposed device to be competitive, the interface roughness between FM2 and TI must be minimized to have $\alpha$ as small as possible (smaller than 0.1). Our proposed device has the lowest energy consumption (Table 1). However, the speed of operation is also lower than the rest, with the energy-delay-product (EDP) being the lowest. For low energy applications, the low energy cost is a higher priority.}
\begin{table}[htp]
\caption{\textcolor{black}{Energy consumption of proposed device compared to existing technologies\cite{tables}. For SRAM and DRAM the 45 nm transistor technology is used. }}
\begin{tabular}{c|c c c}
    Technology & Energy (fJ) & Latency (ns) \\
    \hline\hline
    Proposed device & 10 & 5\\
    \hline
    SRAM & 50 & 2.7\\
    \hline
    DRAM & 65 & 3.4\\
    \hline
    SOT-MRAM & 40 & 2.6\\
\end{tabular}

\label{tab:promise materials}
\end{table}
{\section{Conclusion}}
In this article we proposed an in-memory processing device based on the reciprocal interactions in FM-3DTI heterostructures.  {The main advantage is the integration of the selector-memory units onto a shared transport channel and hence a compact design. Using DFT-calibrated 3D tight binding, NEGF-LLG and Fokker-Planck, we connected the material parameters to switching delay and WER. We showed that a modest ON-OFF suffices for selector action in cache memory, even in presence of current shunting through the sides and bulk. \textcolor{black}{We also showed that as long as the FM1 is not connected to the circuit (no in or out-going current), a modest electric coupling or tunneling between FM1 and 3DTI has only a minor negative effect on the performance. We assumed an in-plane FM2 (type y FM, parallel to spin current polarization) in this work which, to be efficient, needs low damping. This would mean a high-quality interface is crucial. One alternative way is to have FM2 be perpendicular to the spin current polarization, which needs an assisting in-plane magnetic field perpendicular to the spin polarization to break the symmetry. Methods such as adding a second FM layer to use its demagnetization field to act as the assisting magnetic field is shown to work. Other methods, such as structural asymmetry and crystal dependant spin current geometry, have also been proposed\cite{roadmap}. To have SOT switching with 3DTI, a type x ferromagnet, inplane polarization but (anti)parallel to the electric current\cite{fukami_spinorbit_2016} should be more efficient than out-plane FM2 as the magnetic exchange between an out-plane FM2 and 3DTI will lower the efficiency of the current generation by the 3DTI. Improvements in the magnetic exchange between FM1 and 3DTI can make even long-term memory applications possible at a low energy cost. \textcolor{black}{Finally, there are experimental challenges in the electrical switching of a ferromagnet in FM/3DTI heterostructures, and further improvements in the voltage controlled switching of ferromagnet methods are needed.}}\\

\section{Methods}
In the 3D atomistic grid $\{i,j,k\}$, the Hamiltonian looks like \cite{3DTImodel,3DTIhamiltonian,TItight,MTIH}
\begin{eqnarray}
    H &=& H_{3DTI} + H_Z + H_V \label{TI_Ham_eq}\nonumber\\
    H_{3DTI} &=& \sum_{ijk} c^\dagger_{i,j,k}\varepsilon_{3DTI}c_{i,j,k} + \left(c^\dagger_{i,j,k}T_xc_{i+1,j,k}\right. +\nonumber\\ 
           && \left. c^\dagger_{i,j,k}T_yc_{i,j+1,k} + c^\dagger_{i,j,k}T_zc_{i,j,k+1} + h.c.\right)\nonumber\\
    H_Z &=& \sum_{ijk} c^\dagger_{i,j,k}M_0 \mathbf{S}.\mathbf{\sigma}c_{i,j,k}\nonumber\\
    H_V &=& \sum_{ijk}c^\dagger_{i,j,k}V_{g}(x_i,z_k)c_{i,j,k} 
\end{eqnarray}
where the onsite energies $\varepsilon_{3DTI} = (C_0+2C_1+4C_2)I_{4\times4}+(M+2M_1+4M_2) I_{2\times2}\otimes \tau_z$, and the hopping terms $T_{x,y} =-M_2I_{2\times2}\otimes\tau_z - C_2I_{4\times4}+({iA_0}/{2})\sigma_{x,y}\otimes\tau_x$, $T_z =-M_1I_{2\times2}\otimes\tau_z + C_1I_{4\times4}+({iB_0}/{2})\sigma_z\otimes\tau_x$.
For $\mathrm{Bi_2Se_3}$, we use $M = -0.28~ \mathrm{eV}, A_0 = 0.8~ \mathrm{eV}, B_0 = 0.32~ \mathrm{eV}, C_1 = 0.024~ \mathrm{eV}, C_2 = 1.77~ \mathrm{eV}, M_1 = 0.216 ~\mathrm{eV}, M_2 = 2.6~ \mathrm{eV}, C_0 = -0.0083~ \mathrm{eV}$. $\tau$ and $\sigma$ are the Pauli matrices in orbital and spin subspaces respectively while $I$ is the identity matrix. One layer each of the source and drain is included in the Hamiltonian. The fixed FM in the MTJ is not included in the simulation. Our 3D model allows us to separate the bulk and interfacial components of the charge and current densities (Fig.~\ref{fig:Device}).


We employ the Non-Equilibrium Green's Function (NEGF) formalism to analyse electron transport and the overall performance of the device. 
The retarded $G^r$ and  correlation Green's functions $G^n$ are \cite{datta,ghosh}
\begin{eqnarray} 
G^r(E,\mathbf{k_\perp})&=&[EI - H(\mathbf{k_\perp})-\Sigma_S(E,\mathbf{k_\perp})-\Sigma_D(E,\mathbf{k_\perp})]^{-1} \nonumber \\ 
G^n(E,\mathbf{k_\perp})&=& G^r\left(f_S\Gamma_S+f_D\Gamma_D\right)G^{r\dagger} \nonumber\\
\Gamma_{S,D}&=&i\left(\Sigma_{S,D}-\Sigma_{S,D}^\dagger\right)
\end{eqnarray} 
with $f_{S,D}$ the source/drain Fermi-Dirac distributions. 
\textcolor{black}{We assume a uniform cross-section that allows us to Fourier transform the transverse y-z hopping terms into $\vec{k}_\perp$.}  The self-energies $\Sigma_{S,D}$ are calculated recursively for each $E$ and $\vec{k}_\perp$. $\Gamma_{S,D}$ are the corresponding energy broadening matrices. The spin and charge current densities from site $i$ to $j$ and SOT torque are calculated as
\begin{subequations}
\begin{eqnarray}
    \mathbf{J}^{i\rightarrow j}_s &=& \frac{q}{ih} \sum_{\mathbf{k_\perp},E}\mathrm{Tr}\left[\boldsymbol{\sigma}\left(H_{ij}G^n_{ji}-G^n_{ji}H_{ij}\right)\right]\label{spincurrentdensity}\\
    {J}^{i\rightarrow j}_q &=& \frac{q}{ih} \sum_{\mathbf{k_\perp},E}\mathrm{Tr}\left[\left(H_{ij}G^n_{ji}-G^n_{ji}H_{ij}\right)\right]\label{chcurrentdensity}
    \label{currentdensity}
\end{eqnarray}
\end{subequations}
We solve the stochastic LLG equation $(1+\alpha^2)\partial_t \mathbf{m} = \gamma\left(\mathbf{m}\times\mathbf{H}+\alpha\mathbf{m}\times(\mathbf{m}\times\mathbf{H})+\tau_{SOT}\right)$ within a macroscopin model, $\mathbf{H} = \mathbf{H}_{K}+\mathbf{H}_{Th}+\mathbf{H}_D$. $\mathbf{H}_D$ is the demagnetization field. The effective anisotropy field $\mathbf{H}_{K} = ({2\Delta_T k_B T}/{M_s V})\mathbf{(u_0.m)u_0}$, where $\Delta_T$ is the ferromagnet's thermal stability related to anisotropy $K$, $\Delta_T = KV/k_BT$. $\mathbf{u}_0$ is the effective field pointing along $\hat{z}$ and $\hat{y}$ directions for FM1 and FM2 respectively at the start of the simulation.  For thermal fluctuations we add a stochastic field  $\mathbf{H}_{Th} = ({\mathbf{\eta}}/{\mu_0}) \sqrt{{2\alpha k_B T}/{M_{s}\gamma V \Delta t}}  $ \cite{vakili_anatomy_2021}, where $\eta$ is a random vector following a normal distribution with zero average. $\mu_0$ is vacuum permeability, $k_B$ is Boltzmann constant, $T$ is temperature, $M_{s}$ is saturation magnetization, $\gamma$ is the electron gyromagnetic ratio, $V$ and $\Delta t$ are volume of FM2 and simulation time step. 

{\textbf{Acknowledgments.}} We acknowledge useful discussions with Patrick Taylor (ARL), Joe Poon, Md Golam Morshed (UVA) and Supriyo Bandyopadhyay (VCU).
This work is supported by the Army Research Lab (ARL) and in part by the NSF I/UCRC on Multi-functional Integrated System Technology (MIST) Center; IIP-1439644, IIP-1439680, IIP-1738752, IIP-1939009, IIP-1939050, and IIP-1939012. 

\bibliography{bilbography}
\end{document}